\documentclass[aps,prl,preprint,groupedaddress]{revtex4}

\begin{document}


\title{Entropy Bound for the TM Electromagnetic Field in the Half Einstein Universe}


\author{I. Brevik\footnote{Email: iver.h.brevik@ntnu.no}, R. Herikstad and
S. Skriudalen}
\affiliation{Department of Energy and Process Engineering,
Norwegian University of Science and Technology, N-7491 Trondheim,
Norway}


\date{\today}

\begin{abstract}
An explicit calculation is given of the entropy/energy ratio for
the TM modes of the electromagnetic field in the half Einstein
universe. This  geometry provides a mathematically convenient and
physically instructive example of   how the electromagnetic and
thermodynamic quantities behave as a function of the
nondimensional parameter $\delta=1/2\pi aT$, $a$ being the scale
factor and $T$ the temperature. On physical grounds (related to
the relaxation time), it is the case of small $\delta$'s that is
pertinent to thermodynamics. We find that as long as $\delta$ is
small, the entropy/energy ratio behaves in the same way as for the
TE modes. The entropy is thus bounded. The present kind of
formalism makes it convenient to study also the influence from
frequency dispersion. We discuss an example where a sharp cutoff
dispersion relation can in principle truncate the electromagnetic
oscillations in the Einstein cavity such that only the lowest mode
survives.
\end{abstract}


\maketitle

\section{I. Introduction}

Valuable insight can often be obtained by analysing models that
are mathematically simple but nevertheless able to show the
essentials of the physical  properties of a system. The purpose of
the present paper is to present an explicit calculation of
 the energy $E^{TM}$ associated with the transverse
modes (TM) of the electromagnetic field in a spherically symmetric
volume endowed with the metric of the half Einstein universe, and
to combine this with the analogous transverse electric mode result
$E^{TE}$ calculated earlier \cite{brevik02}, in order to obtain
the total electromagnetic energy $E=E^{TM}+E^{TE}$ in the volume.

The calculations will be given for a finite temperature $T$. One
of our motivations is to show, in the case of  the TM modes as
well as in the previous case of TE modes \cite{brevik02}, that the
ratio of  entropy $S^{TM}$ to  thermal energy $E^{TM}$ is limited
in the following sense:
\begin{equation}
\frac{S^{TM}}{E^{TM}}=\frac{4}{3}\beta, \label{1}
\end{equation}
 at high temperatures. Here $\beta=1/T$. (Geometric
units are used.) The half Einstein universe, having received
considerable attention in the past (see, for instance,
Refs.~\cite{kennedy80,bayin93,brevik97,brevik02}), is both
mathematically convenient to handle and provides an instructive
physical example of how the entropy content is limited. The
situation to be considered here thus falls into line with the
current discussion in general about how entropy and energy behave
both quantum mechanically and thermally in conformal field
theories \cite{verlinde00,kutasov01,klemm01}. In particular, a
topic that has attracted particular interest is the Verlinde bound
for the ratio of entropy to energy \cite{verlinde00}. To what
extent the entropy  bound  is realized in physical situations is
usually not clear beforehand; it has  to be investigated by
concrete calculations in each case. Previous calculations on the
entropy bound have usually been done in the high temperature
approximation \cite{kutasov01,klemm01}. We shall also be assuming
that $T$ is high. This is actually a constraint that follows from
thermodynamics: in order for thermal equilibrium to be
established, $T$ has to be much higher than the inverse transit
time for light across the linear dimension of the system. This
point is discussed in more detail in section II.

As mentioned, the case of transverse electric (TE) modes in the
half Einstein universe was considered in Ref.~\cite{brevik02}, and
the bound of Eq.~(\ref{1}) was verified in that case. The
analogous result in the TM case is thus not unexpected, although
 the combined electromagnetic and
thermodynamic behavior is complicated and can hardly be understood
merely by inspection, without calculation. As we shall see, at
higher orders in $\delta$ (see Eq.~(\ref{3}) below), the TM and TE
modes behave somewhat surprisingly, in the sense that the terms
involving $T^3$ and $T^2$ in the energy do not cancel each other.
This contrasts the behaviour known from Casimir theory in flat
space, where the TE and TM modes compensate each other with
respect to the mentioned terms and only the term proportional to
$T$ survives in $E$ to the leading order (cf., for instance,
equation (8.39) in \cite{balian78} or equation (4.44) in
\cite{milton01}). The reason for this behaviour has to be related
to the geometry of the half Einstein universe.

An earlier version of the present paper \cite{brevik05} actually
gave an impetus to the recent investigation of Dowker on general
spacetimes \cite{dowker06}. Dowker's analysis is much more general
than the concrete example on Maxwell fields considered in the
present paper. We think, however,  that there are definite
advantages of going into concrete detail as we are doing here: for
one thing, it becomes easy to analyse the influence from {\it
dispersion} in the wall material. Usually in field theory, one
assumes that the walls have infinity conductivity for all
frequencies. This assumption is unphysical, as there is always a
frequency dispersion present. Of importance in a Casimir context,
is the behaviour of the permittivity $\varepsilon(i\zeta)$ as
function of positive imaginary frequency $\zeta$. For a
dielectric, the susceptibility $\varepsilon(i\zeta)-1$ is in
essence proportional to $(1+\zeta^2/\omega_0^2)^{-1}$, where
$\omega_0$ is the (dominant) resonance frequency. Typically,
$\omega_0 \approx 1.5\times 10^{16}$ rad/s. Thus $\omega_0$ acts
as a soft high-energy cutoff. When the frequencies are much higher
than the resonance frequency one has
$\varepsilon(i\zeta)-1=\omega_p^2/\zeta^2$ for all bodies, metals
or dielectrics, where $\omega_p$ is the plasma frequency. Taking
gold as an example, $\omega_p=1.37\times 10^{16}$ rad/s. For very
high frequencies the photons  do not "see" the metal, and the
number of modes in the cavity is therefore truncated. We shall
return to this point in the final section.

Let us also make some remarks on conformal invariance. The
electromagnetic field is known to be conformally invariant in
$D=4$ spacetime dimensions. Thus the trace of the electromagnetic
energy-momentum tensor is zero when D=4. In higher dimensions the
situation becomes more complicated. The Casimir energy for $D>4$
was first calculated by Ambj{\o}rn and Wolfram \cite{ambjorn83}.
We refer also to the recent analysis of Alnes {\it et al.} on the
electromagnetic field between two parallel hyperplanes in higher
dimensions, considering both metallic and MIT boundary conditions
\cite{alnes06}. Using the axial gauge, the pressure between the
plates was found to be constant, while the energy density was
found to diverge at the boundaries. This peculiar behaviour is a
direct consequence of the lack of conformal invariance when $D>4$.

Another related point worth noticing  is the growing interest in
higher derivative electrodynamics, such as the Lagrangian form
$(F+1/F)$ studied by Novello {\it et al.} \cite{novello04} (here
$F \equiv F_{\mu\nu}F^{\mu\nu})$. One of the motivations of this
kind of approach, among other things, is to model a phase of
cosmic current acceleration. It would be of interest to apply such
a theory also to the half Einstein universe, although we will not
go further with this topic here. The approach has strong
similarities with the $(R+1/R)$ theory in gravity, which has been
thoroughly studied in the recent past (see, for instance,  papers
of Nojiri and Odintsov \cite{nojiri03}).

\section{II. The Half Einstein universe}

 The full static Einstein universe is of general physical interest as
 it is conformally equivalent to all closed FRW metrics. The
 Einstein metric is
 \begin{equation}
 ds^2=-dt^2+a^2(d\chi^2+\sin^2\chi\, d\Omega^2), \label{2}
 \end{equation}
 where $\chi \in [0,\pi]$, $\theta \in [0,\pi]$, and $\phi \in
 [0,2\pi]$. The scale factor $a$ is related to the cosmological
 constant $\Lambda$ through $a=\Lambda^{-1/2}=(4\pi
 G\rho_0)^{-1/2}$, where $\rho_0$ is the energy density of matter
 (dust). The vacuum part of the energy density is $\Lambda/8\pi
 G$; the pressure is $p=-\Lambda/8\pi G$.

 The {\it half} Einstein universe is characterized by $\chi$ being
 restricted to one half of the previous interval, $\chi \in [0,
 \pi/2]$. A general point worth noticing in connection with this universe is that it is
 conformally related to anti-de Sitter space and is of relevance
 to supersymmetry \cite{dowker06}. The universe can be envisioned as
  a three-dimensional spherical
 volume spanned by the 'radius' $\chi$ and the angular coordinates
 $\theta$ and $\phi$, closed by a two-dimensional spherical
 surface at $\chi=\pi/2$. We will take this surface to be
 perfectly conducting, this being an analogy to the Dirichlet
 boundary in the case of a scalar field.

 As there are two dimensional parameters in the problem,  viz.
$\beta$ and $a$, it becomes natural to introduce
 \begin{equation}
 \delta=\frac{\beta}{2\pi a} \label{3}
 \end{equation}
  as a nondimensional parameter. At high temperatures $\delta \ll
  1$, and can  thus be used as  expansion parameter in a
  perturbative analysis. As anticipated  above,
   it is the case of small $\delta$'s that is of
   thermodynamic interest. This can be seen from the following physical argument:
  In order to apply
  thermodynamic formalism to a fluctuating quasi-classical system,
  the temperature has to be sufficiently high to satisfy the
  condition $T \gg 1/\tau$, $\tau$ being the relaxation time \cite{das02}. In
  our case we may take $\tau$ to be of the same order as the
  transit time for light across a distance $a$, {\it i.e.,} $\tau
  \sim a$. [This is essentially the same kind of
  situation as experienced when a narrow beam of light impinges
  upon a liquid surface and makes it bulge outward; cf. the
  classic light pressure experiment of Ashkin and Dziedzic
  \cite{ashkin73} and also the theoretical discussion of it in
  Ref.~\cite{brevik79}. In that case, the relaxation time is of
  the same order as the transit time for {\it sound} to traverse
  the beam.] Thus, in the present case $T\gg 1/a$, which means
  that
  \begin{equation}
  \delta \ll 1 \label{4}
  \end{equation}
  is  the condition for applicability of thermodynamics.

  It is of interest at this point to make a comparison with the
  Casimir effect. Let us assume the usual Casimir configuration,
  namely two parallel metal plates separated by a gap $a$. At
  finite temperature we are dealing with
   discrete Matsubara frequencies, $\zeta_m = 2\pi mT$, with
  $m \geq 0$ an integer. It  turns out that the most important
  contributions to the Casimir force come from the frequency
  region $\zeta_m a \sim 1$, or $m \sim 1/(2\pi aT)$ (cf. the
  discussion on this point in \cite{hoye03}). That is, $m \sim
  \delta$. Then, we may interpret the parameter $\delta$
  physically as the magnitude of the most important Matsubara
  numbers that occur in the analogous Casimir effect. This correspondence
  appears to be physically
  reasonable, though not trivial.

  \section{III. Governing equation. The TM modes}

  We take an orthonormal basis, $ (ad\chi, a\sin \chi d\theta,
  a\sin \chi \sin \theta d\phi)$, and split off the time factor as
  $e^{-i\omega t}$. From Maxwell's equations in curvilinear space
  we can derive the governing equation for $E_\chi$ (or $H_\chi$).
  Denoting collectively these field components by $X$, and writing
  the angular contributions as spherical harmonics, $X(\chi,
  \theta, \phi)=X(\chi)Y_{lm}(\theta, \phi)$, we obtain
  \cite{brevik02}
  \begin{equation}
  \frac{d^2}{d\chi^2}\left(\sin^2 \chi\,X\right)+(\omega
  a)^2\sin^2\chi\,X-l(l+1)X=0. \label{5}
  \end{equation}
  The solution to this equation is
  \begin{equation}
  X \propto \sin^{l-1}\chi \,C_{n-l}^{(l+1)}(\cos \chi), \label{6}
  \end{equation}
  where $n$ is an integer and $C_{n-l}^{(l+1)}$ are the Gegenbauer
  polynomials \cite{abramowitz72} satisfying the differential
  equation
  \begin{equation}
  (1-x^2){C_p^{(\alpha)}}''(x)-(2\alpha
  +1)x\,{{C_p}^{(\alpha)}}'(x)
  +p(p+2\alpha){C_p^{(\alpha)}}(x)=0 \label{7}
  \end{equation}
  for $p \geq 1$. Inserting $X$ into Eq.~(\ref{5}) we obtain the
  eigenfrequencies
  \begin{equation}
  \omega_n=\frac{n+1}{a}, \quad n\geq l \geq 1. \label{8}
  \end{equation}
  To avoid infinities at the origin $\chi=0$, we must have $n-l
  \geq 0$ \cite{brevik97}.

  So far, the boundary conditions have not been considered. The
  transverse magnetic modes are subject to the boundary condition
\begin{equation}
\partial _{\chi}(\sin{\chi}\,{\bf H}_{\perp}) = 0, \hspace{10pt} \chi =
\frac{\pi}{2},\label{9}
\end{equation}
where ${\bf H}_\perp$ is the magnetic field component transverse
to the radius $\chi$. From Maxwell's equations in orthonormal
basis \cite{brevik97} we find, when comparing with the general
solutions of the governing equation (\ref{5}), that the magnitude
$ H_\perp$  of the vector ${\bf H}_\perp$ must be of the form
\begin{equation}
\label{hperp} H_{\perp} \propto
\sin^{l}{\chi}\,C^{(l+1)}_{n-l}(\cos{\chi}). \label{10}
\end{equation}
Thus we have from Eq.~(\ref{9})
\begin{equation}
\partial _{\chi}\left\{\sin^{l+1}{\chi}\,C^{(l+1)}_{n-l}(\cos{\chi})\right\}  =
0 \label{11}
\end{equation}
at the boundary. Observing the recursion relation for the
derivatives of the Gegenbauer polynomials we get
\begin{equation}
\partial_xC^{(\alpha )}_n(0) = (n+2\alpha -1)C^{(\alpha
)}_{n-1}(0)\label{12}
\end{equation}
at the boundary. The condition (\ref{11}) now yields
\begin{equation}
 C^{(l+1)}_{n-l-1} = 0. \label{13}
\end{equation}

As the Gegenbauer polynomials $C^{(\alpha )}_m$ vanish for odd
$m$, $(n-l)$ in Eq.~(\ref{13}) must be even. It follows
that for $n$ even (odd), $l$ must be even (odd) and the
degeneracies are
\begin{equation}
g_n^{(e)} = \sum_{l=2,4,6,...}(2l+1) =
\frac{n-1}{2}(n+2),\hspace{10pt}n
\hspace{10pt}\rm{even},\label{14}
\end{equation}
\begin{equation}
g_n^{(o)} = \sum_{l=1,3,5,...}(2l + 1) =
\frac{n}{2}(n+1),\hspace{10pt}n \hspace{10pt}\rm{odd}.\label{15}
\end{equation}
This leads to the following expression for the logarithm of the
partition function $Z^{TM}$ for the TM modes:
\begin{eqnarray}
\nonumber \ln{Z^{\rm{TM}}} = - \sum_{n=1}^\infty (2n-1)(n+1)\ln{[1-e^{-2\pi(2n+1)\delta}]}  \nonumber \\
 - \sum_{n=1}^\infty(2n-1)n\ln{[1-e^{-4\pi n \delta }]}. \label{16}
\end{eqnarray}
From this we identify two types of sums:
\begin{equation}
 G^{(\alpha )} =
\sum_{n=1}^{\infty}n^{\alpha}\ln{[1-e^{-2\pi (2n+1)\delta}]},
\hspace{10pt}\alpha = 0,1,2, \label{17}
\end{equation}
and
\begin{equation}
 H^{(\alpha )} =
\sum_{n=1}^{\infty}(2n-1)^{\alpha}\ln{[1-e^{-4\pi n\delta}]},
\hspace{10pt}\alpha = 1,2, \label{18}
\end{equation}
so that
\begin{equation}
 \ln{Z^{\rm{TM}}} = -2G^{(2)} - G^{(1)} +
G^{(0)} - \frac{1}{2}(H^{(2)} + H^{(1)}). \label{19}
\end{equation}
We expand the logarithm in Eq.~(\ref{17}) and take the derivative
with respect to $\delta$:
\begin{eqnarray}
\frac{\partial G^{(\alpha )}}{\partial \delta} =
2\pi\sum_{n=1}^{\infty}n^{\alpha}(2n+1)\sum_{k=1}^{\infty}e^{-2\pi
(2n+1)k\delta} \nonumber \\
= \frac{1}{i}\int_C \rm{d}s(2\pi \delta)^{-s}\,\Gamma (s)\zeta
(s)\sum_{n=1}^{\infty}n^{\alpha}(2n+1)^{1-s}, \label{20}
\end{eqnarray}
$\zeta(s)$ being Riemann's zeta function. In the last step above
we made use of the relation
\begin{equation}
e^{-x}=\frac{1}{2\pi i}\int_C ds\,x^{-s}\,\Gamma(s), \label{21}
\end{equation}
where the integration contour is a line parallel to the imaginary
axis at a sufficiently large value of $\Re s$.

As we need to work out the sum over $n$  it will prove worthwhile
to derive a general expression for the integral in Eq.~(\ref{20}).
To this end we change the summation variable,
\begin{equation}
\sum_{n=1}^{\infty}n^{\alpha}(2n+1)^{1-s} =
\sum_{k=3,5,...}^\infty \left (\frac{k-1}{2} \right
)^{\alpha}k^{1-s}. \label{22}
\end{equation}
The terms in brackets can be expressed as a binomial series. We
add and subtract the $k=1$ term as well as the even terms, and
insert the result into Eq.~(\ref{20}) to get

\begin{eqnarray}
 \frac{\partial G^{(\alpha )}}{\partial \delta} =
\frac{1}{2^{\alpha}i}\int_C \rm{d}s(2\pi \delta)^{-s}\,\Gamma
(s)\zeta (s) \nonumber \\
\times \sum_{l=0}^{\alpha} \left( \begin{array}{c} \alpha
\\l
\end{array} \right)
(-1)^{l}\Big\{(1-2^{1-s+\alpha -l})\zeta (s-1-\alpha + l) -
1\Big\}.\label{23}
\end{eqnarray}
When $\alpha = 0$ this expression becomes
\begin{equation}
\frac{\partial G^{(0)}}{\partial \delta} = \frac{1}{i}\int_C
\rm{d}s(2\pi \delta)^{-s}\Gamma (s)\zeta (s)\left\{
(1-2^{1-s})\zeta (s-1) - 1 \right\}.\label{24}
\end{equation}
The two terms in the integrand have poles for $s=0,2$ and $s=0,1$
respectively. There are thus three poles in all.

When $\alpha =1$ we have
\begin{equation}
\frac{\partial G^{(1)}}{\partial \delta} = \frac{1}{2i}\int_C
\rm{d}s(2\pi \delta)^{-s}\Gamma (s)\zeta (s)\left\{
(1-2^{2-s})\zeta (s-2) - (1-2^{1-s})\zeta (s-1)\right\}.\label{25}
\end{equation}
The poles are at $s=1,3$ in the first term, and at $s=0,2$ in the
last term.

 The final integral is for $\alpha = 2$,
\begin{eqnarray}
\frac{\partial G^{(2)}}{\partial \delta} &=& \frac{1}{4i}\int_C \rm{d}s(2\pi \delta)^{-s}\,\Gamma (s)\zeta (s)  \\
\nonumber && \times \left\{(1-2^{3-s})\zeta (s-3) -
2(1-2^{2-s})\zeta (s-2) + (1-2^{1-s})\zeta
(s-1)\right\},\label{26}
\end{eqnarray}
with poles at $s=0, 4$, $s=1,3$ and $s=0,2$ in the three terms
respectively. Calculating all residues and collecting terms we
find
\begin{equation}
\frac{\partial G^{(0)}}{\partial \delta} =  \frac{\pi}{24\delta
^2} - \frac{1}{\delta } + \frac{11\pi}{12}, \label{27}
\end{equation}
\begin{equation}
\frac{\partial G^{(1)}}{\partial \delta}  =  \frac{\zeta (3)}{8\pi
^2 \delta ^3}
 - \frac{\pi}{48\delta ^2} + \frac{1}{24\delta } + \frac{\pi}{24}, \label{28}
 \end{equation}
 \begin{equation}
\frac{\partial G^{(2)}}{\partial \delta}  = \frac{\pi}{960\delta
^4} - \frac{\zeta (3)}{8\pi ^2 \delta ^3} + \frac{\pi}{96\delta
^2} - \frac{1}{24\delta } - \frac{\pi}{160}. \label{29}
\end{equation}

We turn now to $H^{(\alpha)}$, following the same steps as for
$G^{(\alpha )}$. First, we express the derivative in the form of
an  integral,
\begin{equation}
\frac{\partial H^{(\alpha)}}{\partial \delta}=\frac{2}{i}\int_C
{\rm d}s(4\pi \delta)^{-s}\Gamma(s)\zeta(s)\sum_{n=1}^\infty
n^{1-s}(2n-1)^\alpha. \label{30}
\end{equation}
Again using the binomial series in the $n$ sum we arrive at the
following generic expression:
\begin{eqnarray}
 \frac{\partial H^{(\alpha )}}{\partial \delta} =
\frac{2}{i}\int_C\rm{d}s(4\pi\delta)^{-s}\Gamma(s)\zeta (s)
\nonumber \\
\times \sum_{l=0}^{\alpha}\left( \begin{array}{c} \alpha
\\l
\end{array} \right)
(-1)^l 2^{\alpha - l}\zeta (s-1-\alpha +l).\label{31}
\end{eqnarray}
For $\alpha =1$ we here get
\begin{equation}
\frac{\partial H^{(1)}}{\partial \delta} =
\frac{2}{i}\int_C\rm{d}s(4\pi\delta)^{-s}\, \Gamma
(s)\zeta(s)\left\{ 2\zeta (s-2) - \zeta (s-1)\right\},\label{32}
\end{equation}
with poles at $s=1,3$ in the first term and $s=0,1,2$ in the
second term. Similarly, for $\alpha = 2$ we get
\begin{equation}
\frac{\partial H^{(2)}}{\partial \delta} =
\frac{2}{i}\int_C\rm{d}s(4\pi\delta)^{-s} \,\Gamma (s)\zeta(s)
\left\{ 4\zeta (s-3)-4\zeta (s-2) + \zeta(s-1) \right\},
\label{33}
\end{equation}
with poles at $s=0,4$, $s=1,3$ and $s=0,1,2$ in the first, second
and third term respectively. Calculating all residues we obtain
\begin{equation}
\frac{\partial H^{(1)}}{\partial \delta}=\frac{\zeta (3)}{4\pi
^2\delta ^3} -\frac{\pi}{24\delta ^2} + \frac{1}{3 \delta } -
\frac{\pi}{6}, \label{34}
\end{equation}
\begin{equation}
\frac{\partial H^{(2)}}{\partial \delta}=\frac{\pi}{240\delta ^4}
- \frac{\zeta (3)}{2\pi ^2 \delta ^3} + \frac{\pi}{24\delta ^2} -
\frac{1}{6\delta} + \frac{\pi}{10}. \label{35}
\end{equation}
Inserting the various terms into Eq.~(\ref{19}) we obtain the
following expression for the energy $E^{TM}=-\partial/\partial
\beta \ln Z^{TM}$ of the TM modes:
\begin{equation}
2\pi a E^{\rm{TM}} = \frac{\pi}{240 \delta ^4} - \frac{\zeta
(3)}{4\pi ^2\delta ^3} - \frac{\pi}{24\delta ^2} +
\frac{25}{24\delta} - \frac{221\pi}{240}. \label{36}
\end{equation}
We similarly calculate the free energy $F^{TM}=-(1/\beta)\ln
Z^{TM}$:
\begin{equation}
\beta F^{TM} = -\frac{\pi}{720\delta ^3} +\frac{\zeta (3)}{8\pi
^2\delta ^2} + \frac{\pi}{24\delta} + \frac{25}{24}\ln\delta -
\frac{221\pi}{240}\delta, \label{37}
\end{equation}
and finally the entropy $S^{TM}=\beta^2\partial F^{TM}/\partial
\beta$:
\begin{equation}
S^{TM} = \frac{\pi}{180\delta ^3} - \frac{3\zeta (3)}{8\pi
^2\delta^2} - \frac{\pi}{12\delta} - \frac{25}{24}\ln\delta +
\frac{25}{24}. \label{38}
\end{equation}
To leading order in $\delta$ this yields
\begin{equation}
\frac{S^{TM}}{E^{TM}}=\frac{4}{3}\beta, \quad \delta \ll 1.
\label{39} \end{equation} Thus, in the high temperature limit the
entropy of the TM modes is bounded, just as for the TE modes; cf.
Eq.~(\ref{1}).

\section{IV. Comparison with the TE modes. The total field quantities}

The equality of the entropy/energy ratios for the TM and TE modes
at high temperatures - meaning physically, as we have seen - that
the temperature $T$ is much higher than the inverse relaxation
time $1/\tau$ - is as we might expect in view of the separability
of the TM and TE modes in spherical geometry (cf., for instance,
Ref.~\cite{milton04}). But according to our calculations there are
differences between these modes as regards higher order terms in
$\delta=\beta/2\pi a$. The  expressions pertaining to the TE modes
are
\begin{equation}
2\pi a E^{\rm{TE}}  =
\frac{\pi}{240\delta^{4}}-\frac{\zeta(3)}{4\pi^{2}\delta^{3}}-\frac{\pi}{24\delta^{2}}+\frac{13}{24\delta}-\frac{41\pi}{240},\label{40}
\end{equation}
\begin{equation}
\label{freeEnValue} \beta
F^{\rm{TE}}=-\frac{\pi}{720\delta^{3}}+\frac{\zeta(3)}{8\pi^{2}\delta^{2}}+\frac{\pi}{24\delta}+\frac{13}{24}\ln\delta-\frac{41\pi}{240}\delta,\label{41}
\end{equation}
\begin{equation}
\label{entropyvalue} S^{\rm
TE}=\frac{\pi}{180\delta^{3}}-\frac{3\zeta (3)}{8 \pi ^{2}
\delta^{2}}-\frac{\pi}{12 \delta}-\frac{13}{24}\ln
\delta+\frac{13}{24}.\label{42}
\end{equation}
These results were obtained in \cite{brevik02}  via the same
method as above, and also, as an independent check, via use of the
Euler-Maclaurin sum formula. Adding the contributions from the TM
and TE modes we obtain the following total field quantities, when
expressed in conventional units,
\begin{equation}
E=\frac{\pi^4}{15}a^3T^4-2\zeta(3)a^2T^3-\frac{\pi^2}{6}aT^2+\frac{19}{12}T-\frac{131}{240a},
\label{43}
\end{equation}
\begin{equation}
F=-\frac{\pi^4}{45}a^3T^4+\zeta(3)a^2
T^3+\frac{\pi^2}{6}aT^2-\frac{19}{12}\,T\ln(2\pi
aT)-\frac{131}{240\pi a}, \label{44}
\end{equation}
\begin{equation}
S=\frac{4\pi^4}{45}a^3T^3-3\zeta(3)a^2T^2-\frac{\pi^2}{3}aT+\frac{19}{12}\ln(2\pi
aT)+\frac{19}{12}. \label{45}
\end{equation}
Surprisingly enough, the TE and TM contributions to the $T^3$ and
$T^2$ terms in the expression (\ref{43}) for $E$ do not cancel
out.

\section{Summary and further discussion}

Let us first summarize a couple of points:

$\bullet$ We have assumed the parameter $\delta=\beta/2\pi
a=1/(2\pi aT)$ to be small. This is in accordance with the
requirement of classical thermodynamics: under  equilibrium
conditions $T$ has to be much larger than the inverse relaxation
time \cite{das02}. When comparing with the Casimir effect between
two metal plates, $\delta$ may be given a physical interpretation
as the magnitude of the most dominant Matsubara numbers
\cite{hoye03}.

$\bullet$ The most striking result of the above calculation is
that the TE and TM modes do not compensate each other to orders
$T^3$ and $T^2$ in the expression (\ref{43}) for the total energy.
One might expect beforehand that the mentioned compensation should
take place here as well as in the known case of flat space,
considered earlier in connection with the Casimir effect
\cite{balian78,milton01}. We attribute the non-compensation to the
properties of the Einstein metric. The formalism is generally too
complicated to be  transparent beforehand (it may be noted here
that the degeneracies of the TM modes as given in Eqs.~(\ref{14})
and (\ref{15}) are complementary to those holding for the TE modes
\cite{brevik02}).

$\bullet$ We shall consider below some aspects related to
dispersion. As a preliminary step, let us give first a brief
account of the essence of the formalism for calculating the total
energy associated with the individual TE and TM  field oscillation
modes, characterized by the numbers $n$ and $l$ \cite{brevik97}.

Consider first the TE modes. As in \cite{brevik97} it is
convenient to change the meaning of $n$, such that $n$ runs from 0
upwards. The eigenfrequencies can then be expressed as
\begin{equation}
 \omega_n^{TE}=\frac{2n+l+2}{a}, \label{46}
\end{equation}
with $l=1,2,3...$ as before. The "radial" magnetic field component
can be written as
\begin{equation}
H_\chi=A^{TE}l(l+1)\sin^{l-1}\chi \,C_{2n+1}^{(l+1)}(\cos
\chi)\,Y_{lm}(\theta, \phi), \label{47}
\end{equation}
where $A^{TE}$ is a normalization constant. The other magnetic
field components are, in an orthonormal basis,
\begin{equation}
H_\theta=\frac{A^{TE}}{\sin
\chi}\frac{d}{d\chi}\left[\sin^{l+1}\chi \, C_{2n+1}^{(l+1)}(\cos
\chi)\right] \partial_\theta Y_{lm}, \label{48}
\end{equation}
\begin{equation}
H_\phi=\frac{imA^{TE}}{\sin \chi}\frac{d}{d\chi}\left[
\sin^{l+1}\chi\,C_{2n+1}^{(l+1)}(\cos\chi)\right]\frac{Y_{lm}}{\sin\theta}.
\label{49}
\end{equation}
The time factor $\exp(-i\omega t)$ is assumed everywhere. The
electric field components $E_\theta$ and $E_\phi$, not given here,
follow from Maxwell's equations. Using these field expressions, we
obtain by integrating over the volume the following expression for
the total energy:
\begin{equation}
E_{nl}^{TE}=\frac{\pi}{8}a^3
|A^{TE}|^2\,l(l+1)(2n+l+2)\,\frac{2^{-2l}\Gamma(2n+2l+3)}{(2n+1)!\,[\Gamma(l+1)]^2}.
\label{50}
\end{equation}
As for the TM modes, we write analogously
\begin{equation}
\omega_n^{TM}=\frac{2n+l+1}{a}, \label{51}
\end{equation}
with $n=0,1,2,...$ The electric field components can be written as
\begin{equation}
E_\chi=A^{TM}l(l+1)\sin^{l-1}\chi\,C_{2n}^{(l+1)}(\cos
\chi)\,Y_{lm}, \label{52}
\end{equation}
\begin{equation}
E_\theta= \frac{A^{TM}}{\sin\chi}\frac{d}{d\chi}\left[
\sin^{l+1}\chi\,C_{2n}^{(l+1)}(\cos \chi)\right] \partial_\theta
Y_{lm}, \label{53}
\end{equation}
\begin{equation}
E_\phi= \frac{imA^{TM}}{\sin\chi}\frac{d}{d\chi}\left[
\sin^{l+1}\chi\,C_{2n}^{(l+1)}(\cos
\chi)\right]\frac{Y_{lm}}{\sin\theta}, \label{54}
\end{equation}
with corresponding expressions for the transverse components
$H_\theta$ and $H_\phi$. The total energy becomes in this case
\begin{equation}
E_{nl}^{TM}=\frac{\pi}{8}a^3|A^{TM}|^2l(l+1)(2n+l+1)\,\frac{2^{-2l}\,\Gamma(2n+2l+2)}{(2n)!\,[\Gamma(l+1)]^2}.
\label{55}
\end{equation}
$\bullet$ We are now able to discuss how frequency dispersion
restricts the modes of oscillations in the cavity. As for
dispersion relation, we may for a dielectric take a Lorentz (or
Sellmeir) form as mentioned already in the Introduction,
\begin{equation}
\varepsilon(i\zeta)=1+\frac{\varepsilon(0)-1}{1+\zeta^2/\omega_0^2},
\label{56}
\end{equation}
where $\zeta$ is the imaginary frequency and $\omega_0$ the
resonance frequency, the latter being a  soft frequency cutoff.
However, there are no thermodynamic restrictions preventing us
from assuming that there is a simple sharp cutoff at
$\zeta=\omega_0$, so let us adopt this simple prescription.
Moreover, when assuming an ideal metal, we have that
$\varepsilon(0)=\infty$. Our dispersion model becomes accordingly
\begin{equation}
\varepsilon(i\zeta)=\left\{ \begin{array}{ll} \infty, &
\zeta \leq \omega_0 \\
1                                                          , &
\zeta
> \omega_0.
\end{array}
\right. \label{57}
\end{equation}
As for resonance frequency $\omega_0$, we shall take the same
typical value as mentioned earlier,
\begin{equation}
\omega_0=1.5\times 10^{16}\,{\rm rad/s}. \label{58}
\end{equation}
Consider now the TE modes, where the eigenfrequencies are given in
dimensional units as $\omega_n^{TE}=(c/a)(2n+l+2)$. The lowest
mode is obtained for $n=0, l=1$ as $(\omega_0^{TE})_{min}=3c/a$.
Let us choose the "radius" of the cavity to be small,
\begin{equation}
a=45\,{\rm nm} \label{59}
\end{equation}
(this radius is large enough to permit use of macroscopic
electromagnetic theory in the material). Then,
$(\omega_0^{TE})_{min}=2\times 10^{16}$ rad/s. This mode can
according to (\ref{58}) {\it not} exist in the cavity; the
permittivity in the walls is simply equal to one.

In the TM case, we obtain analogously from
$\omega_n^{TM}=(c/a)(2n+l+1)$ that $(\omega_0^{TM})_{min}=2c/a$
for $n=0, l=1$.  This  is the lowest possible oscillation mode in
the cavity (if we assume that there is no restriction coming from
dispersion at all). Numerically, $(\omega_0^{TM})_{min}=1.33\times
10^{16}$ rad/s. This oscillation mode can thus exist also in the
present case, under the given conditions. Our choice of values in
Eqs.~(\ref{58}) and (\ref{59}) has thus managed to make the lowest
possible mode in the cavity to be the only real one. Our example
is somewhat extreme, but it serves to demonstrate the important
effects of dispersion.

Equations (\ref{50}) and (\ref{55}) permit us to calculate the
field energy in each mode. In the present case only (\ref{55}) is
actual, and it gives for the lowest mode
\begin{equation}
E_{01}^{TM}=\frac{3\pi}{4}a^3 |A^{TM}|^2. \label{60}
\end{equation}

\bigskip

{\bf Acknowledgments}

\bigskip

We thank Sergei Odintsov and Stuart Dowker for valuable remarks
and discussions.


\end{document}